\documentstyle[11pt,newpasp,twoside,epsf]{article}
\newfont{\sfsl}{cmssqi8 scaled 1300}
\newcommand{\gcs}{{\sfsl HIFLUGCS}}

\def\edcomment#1{\iffalse\marginpar{\raggedright\sl#1\/}\else\relax\fi}
\reversemarginpar

\begin{document}
\title{On the Determination of the Mean Cosmic Matter Density and
the Amplitude of Density Fluctuations}
\author{Thomas H. Reiprich}
\affil{Department of Astronomy,
University of Virginia,
PO Box 3818,
Charlottesville, VA 22903-0818,
USA; thomas@reiprich.net
}
\begin{abstract}
The cosmological implications from a new estimate of the local X-ray galaxy
cluster abundance are summarized. The results are then compared to independent
observations. It is suggested that `low' values for the mean cosmic matter
density and the amplitude of mass density fluctuations currently do not appear
unreasonable observationally.
\end{abstract}
%
%\vspace{-0.6cm}
\section{Constraints from the \gcs\ Mass Function}
%
%Unless stated otherwise,
%$\Omega_{\rm m}=1, \Omega_{\Lambda}=0, H_0=50\,h_{50}^{-1}\,\rm km/s/Mpc$ with
%$h_{50}=1$ is used throughout.
A new X-ray selected and X-ray flux-limited galaxy cluster sample has been
constructed (\gcs, the HIghest X-ray FLUx Galaxy Cluster Sample, Reiprich \&
B\"ohringer 2002). Based on the
ROSAT All-Sky Survey the 63 brightest clusters
with galactic latitude $\vert b_{\rm II} \vert \geq 20$\,deg and flux
$f_{\rm X}(0.1-2.4\,{\rm keV})\ge
2\times10^{-11}\,{\rm ergs\,s^{-1}\,cm^{-2}}$ have been compiled. Gravitational
masses have been determined 
utilizing intracluster gas density profiles, derived mainly from ROSAT PSPC
pointed observations, and gas temperatures, as published mainly from
ASCA observations, assuming hydrostatic equilibrium.
This sample and an extended sample of 106 galaxy clusters has been used to
establish the X-ray luminosity--gravitational mass relation.

From the complete sample and the individually determined masses the galaxy
cluster mass function has been determined and used to constrain the mean cosmic
matter density and the amplitude of mass density fluctuations.
Comparison to Press--Schechter type model mass functions in the framework
of cold dark matter cosmological models and a Harrison-Zeldovich initial
density fluctuation spectrum yields the constraints
$\Omega_{\rm m}=0.12^{+0.06}_{-0.04}$ and
$\sigma_8=0.96^{+0.15}_{-0.12}$ (90\% c.l.). The degeneracy between
$\Omega_{\rm m}$ and $\sigma_8$ previously encountered for local cluster samples
therefore has been broken mainly due to the large covered mass range (Fig.~1;
see section 5.3.2 in Reiprich \& B\"ohringer 2002 for more details).
Various possible systematic uncertainties have been quantified.
Adding all identified systematic uncertainties to the
statistical uncertainty in a worst case fashion results in an upper limit
$\Omega_{\rm m}<0.31$. For comparison to previous results a relation
$\sigma_8=0.43\,\Omega_{\rm m}^{-0.38}$ has been derived.

Two further constraints on $\Omega_{\rm m}$ obtained from the \gcs\ clusters
agree well with the above results. The mean intracluster gas fraction combined
with independent estimates of the baryon density yields the upper limit
$\Omega_{\rm m}\la 0.34$.
Calculation of the median mass-to-light ratio for 18 clusters in common to the
sample of Girardi et al.\ (2000) combined with estimates of the total luminosity
density in the Universe yields $\Omega_{\rm m}\approx 0.15$ (Reiprich 2001).

The mass function has been integrated to show that the contribution of mass bound
within virialized cluster regions to the total matter density is small; i.e.,
$\Omega_{\rm cluster}=0.012^{+0.003}_{-0.004}$ for cluster masses larger than
$6.4^{+0.7}_{-0.6}\times 10^{13}\,h_{50}^{-1}\,M_{\odot}$. If light traces mass
this also implies that most galaxies sit outside clusters.
\begin{figure}
%\vspace{-0.6cm}
\plottwo{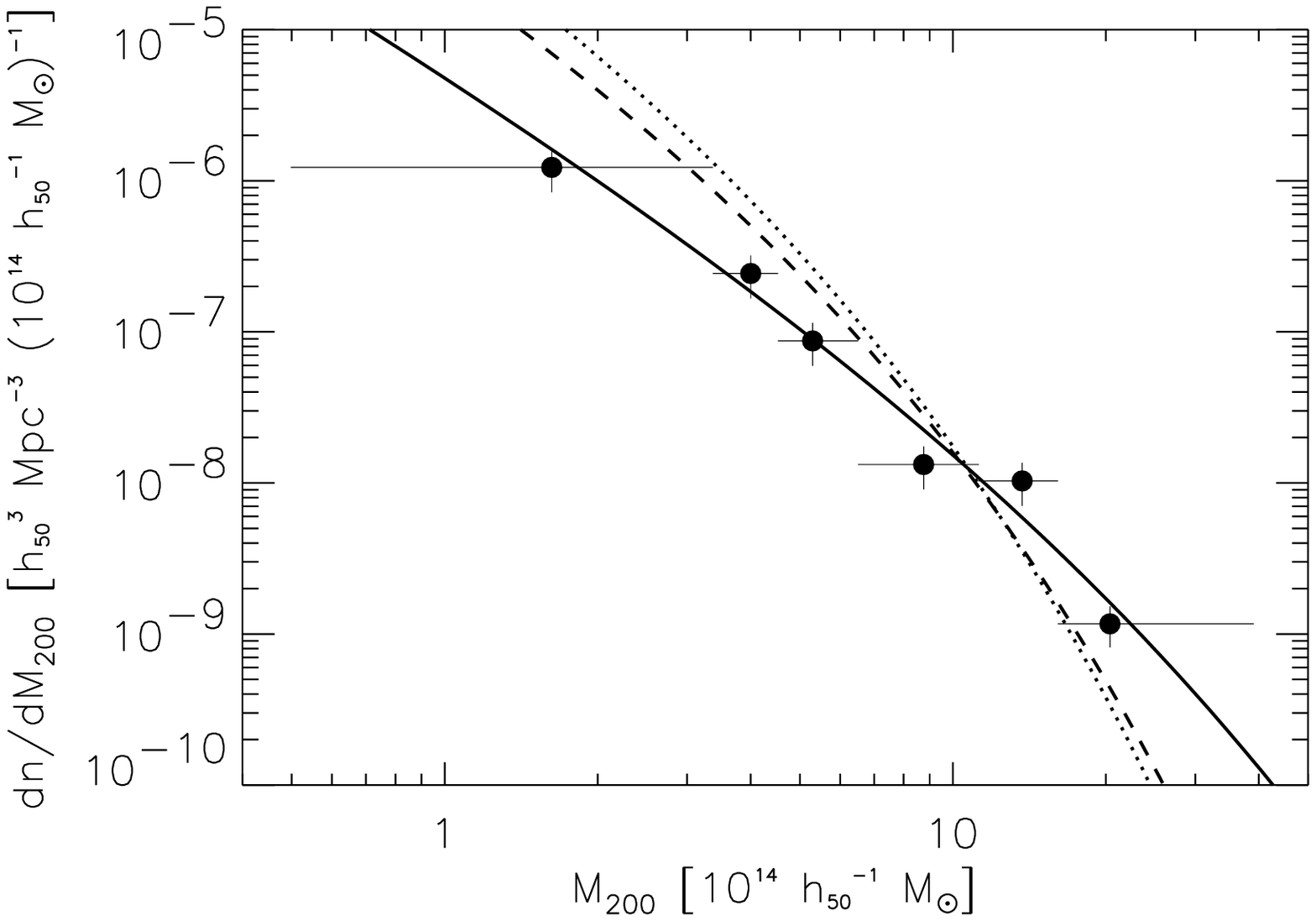}{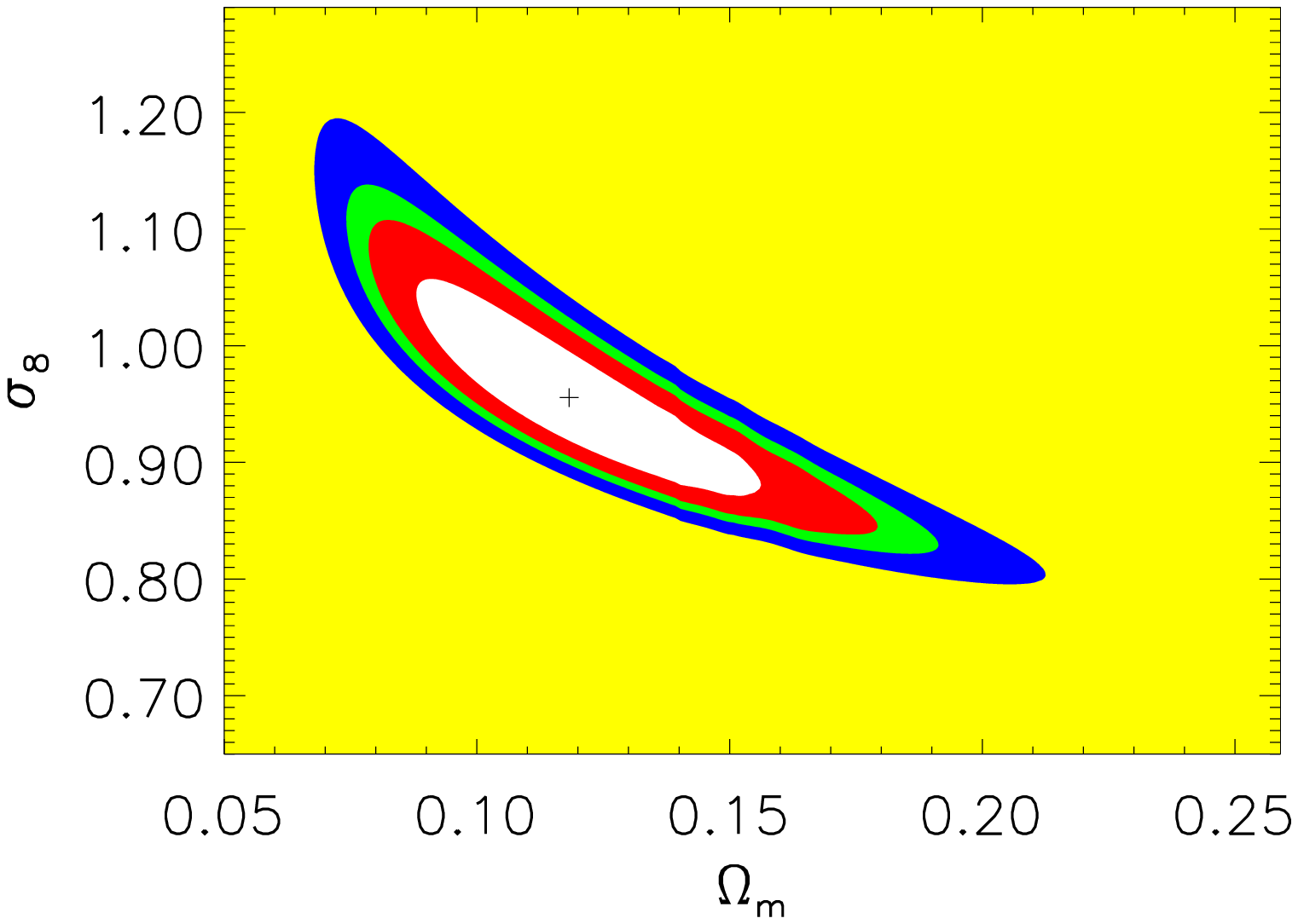}
%\vspace{-0.3cm}
\caption{{\bf Left:} \gcs\ differential mass function compared to the best fit model mass
function with $\Omega_{\rm m}=0.12$ and $\sigma_8=0.96$ (solid line). Also shown
are the best fit model mass functions for fixed $\Omega_{\rm m}=0.5$
($\Rightarrow\sigma_8=0.60$; dashed line) and $\Omega_{\rm m}=1.0$
($\Rightarrow\sigma_8=0.46$; dotted line). {\bf Right:} Statistical confidence
contours. The cross indicates the minimum and ellipses the 68\%, 90\%, 95\%, and
99\% c.l.\ for two interesting parameters.}
%\vspace{-0.48cm}
\end{figure}
\section{Comparison to Independent Observations}
\begin{figure}
%\vspace{-0.6cm}
\plottwo{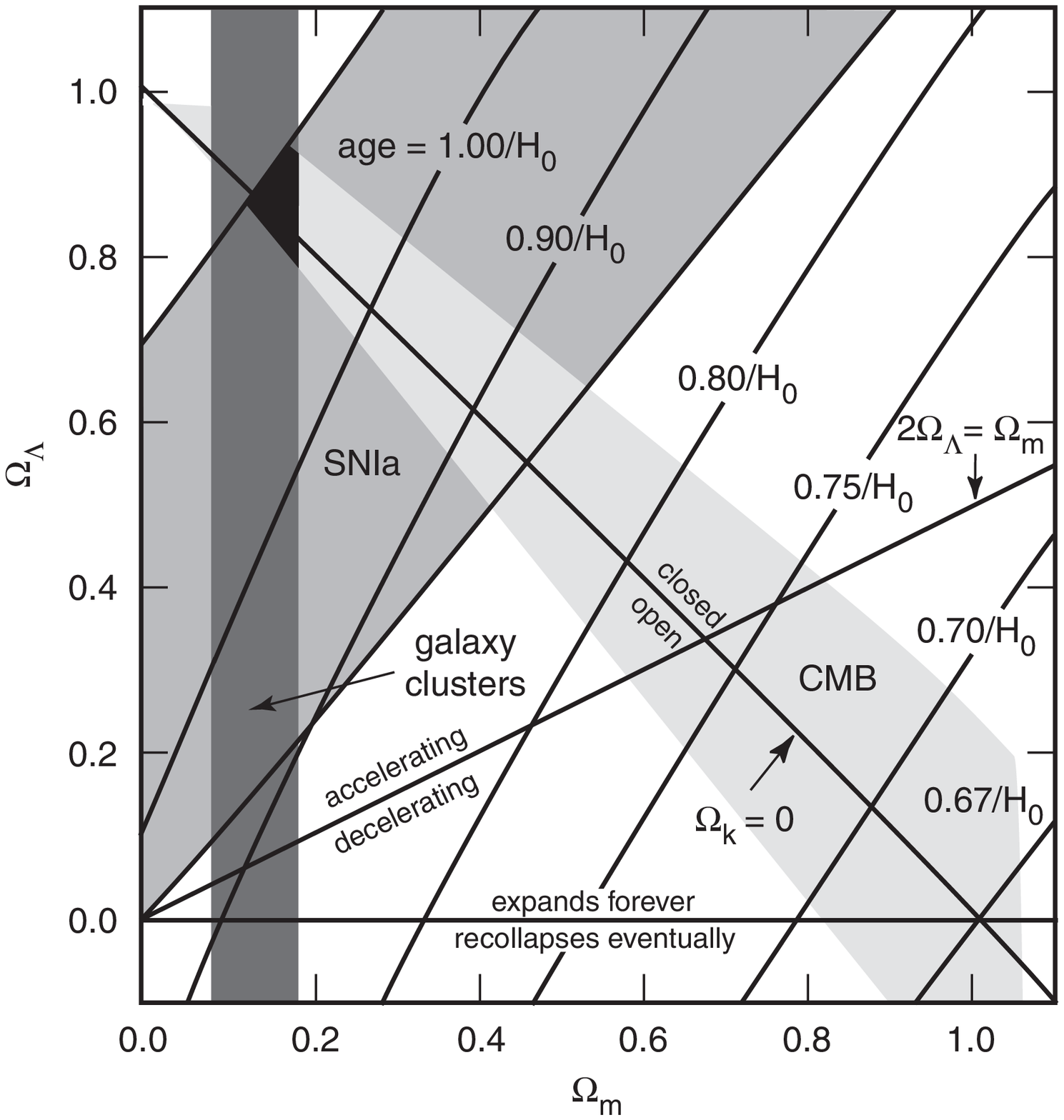}{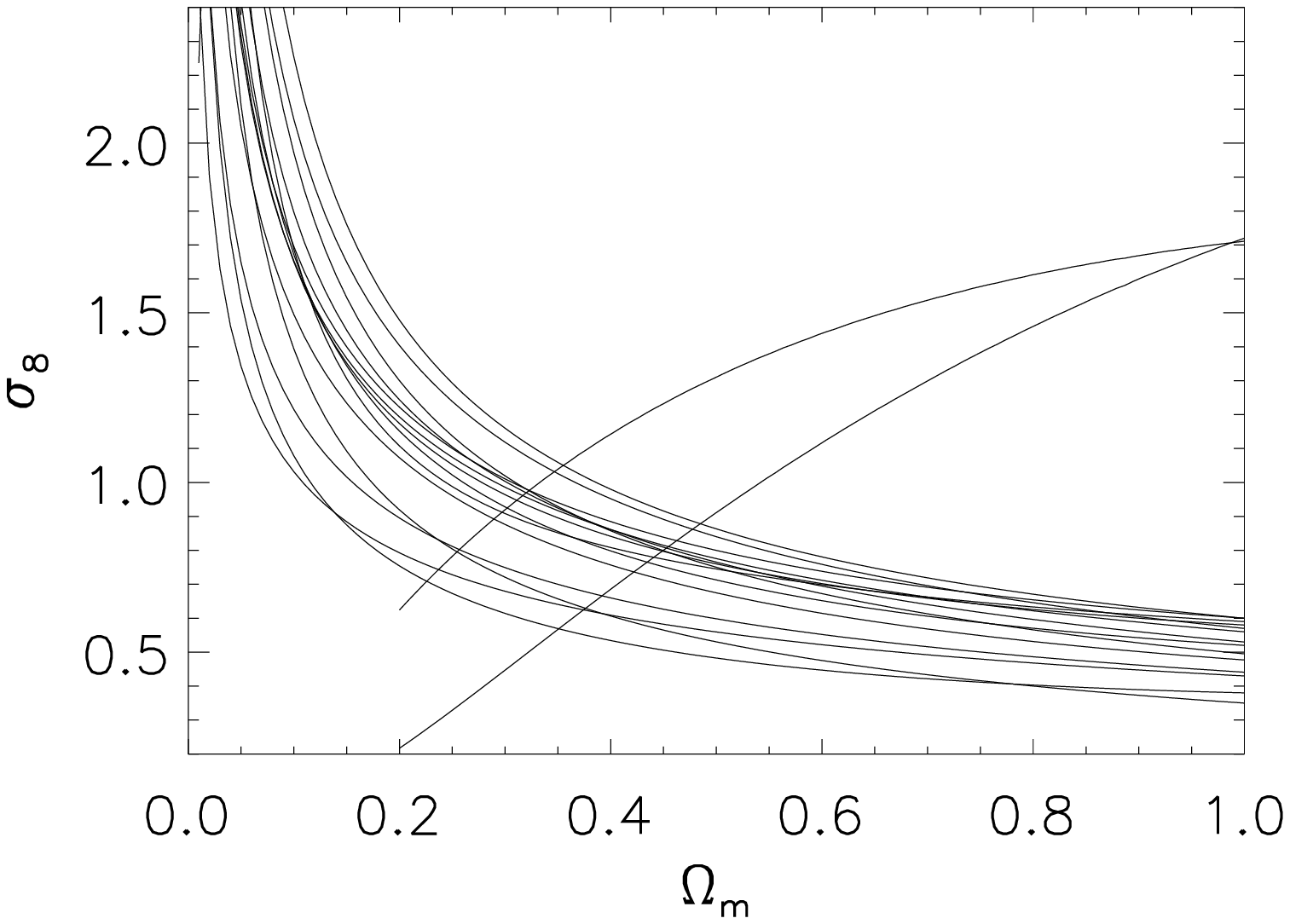}
%\vspace{-0.3cm}
\caption{{\bf Left:} Statistical constraints from the \gcs\ mass function
compared to constraints from supernovae (Perlmutter et al.\ (1999) and cosmic
microwave background (de Bernardis et al.\ 2000) measurements. Note that the worst
case upper limit for \gcs\ including all identified systematic uncertainties is
given by $\Omega_{\rm m}<0.31$. Basic figure
adopted from de Bernardis et al.\ (2000). {\bf Right:} Comparison of $\Omega_{\rm
m}$--$\sigma_8$ relations obtained from cluster abundances. See text for
details. The two $\sim$orthogonal lines indicate the COBE normalization as taken
from Bunn \& White (1997). The upper line is calculated for a flat and the lower
line for an open Universe.}
%\vspace{-0.48cm}
\end{figure}
The constraints from Sect.~1 are compared to independent constraints
in Fig.~2. Shown are (95\%) confidence contours from measurements of distant
type Ia supernovae and temperature fluctuations in the cosmic microwave
background. These confidence contours overlap with the 90\% statistical
uncertainty region from the mass function and the overall region of consistency
indicates that the expansion of the (apparently flat) Universe is currently
accelerating. 

Also shown are $\Omega_{\rm m}$--$\sigma_8$ relations as determined from cluster
abundance measurements compiled from the examples given in Reiprich \&
B\"ohringer (2002) plus three more recent estimates from Seljak (2002), Viana et
al.\ (2002), and Bahcall et al.\ (2002), as well as the `COBE normalization'.
The various inherent differences of the individual analyses
which affect the shape of these lines, e.g., the value of the primordial power
spectral index, $n$, and the calculation of the transfer
function, will be ignored in the following discussion.
Note also that several other methods exist
to estimate $\sigma_8$. A discussion of them is not possible here but
it appears fair to state that the range of other estimates of $\sigma_8$ is
similar to the range of the cluster results shown here, at least for the
relevant $\Omega_{\rm m}$ range. 

The four lowest cluster normalizations (all published in
2002) do not fit the higher `canonical cluster normalization' and recently the
categorization `low' and `high' cluster normalization appears to have been
adopted. If the low cluster normalization turns out to be close to
the `true' value and the Universe is flat then comparison to the COBE results
indicates a low value for $\Omega_{\rm m}$ (0.20--0.25). Should independent
measurements fix $\Omega_{\rm m}$ to a value very close to the `bandwagon' (J.P.
Henry) value 0.3 then this could be an exciting indication for the
presence of inflationary gravitational waves biasing upwards the COBE
normalization.

Attention on the reasons for the discrepant cluster results has been focussed
mainly on different normalizations of cluster mass--X-ray temperature relations.
However, this is only one source of uncertainty as is immediately obvious by
noting that, e.g., several determinations relied entirely on optical data (i.e.,
optically selected cluster samples and masses estimated from cluster richness or
galaxy velocity dispersion). Other sources of uncertainty include the following.
The different cluster samples vary in the covered volume size and are
therefore affected by cosmic variance in different ways. Slight systematic
differences between optical and X-ray cluster selection and mass estimates exist
as discussed, e.g.,
in Reiprich (2001) and Reiprich \& B\"ohringer (2002). Different methods to
determine an overall cluster X-ray temperature yield systematically different
results. It is not clear whether the slope of the mass--temperature
relation is correctly described by $M\propto T^{1.5}$ over the entire mass
range. The treatment of observed scatter in relations varies.
Different samples may contain different fractions of clusters undergoing
a major merger. Optical and X-ray mass estimates of merging clusters tend to
overestimate their mass and may therefore lead to overestimates of $\sigma_8$
(e.g., Randall et al.\ 2002).

Nevertheless the uncertainty in the normalization of the $M$--$T$
relation is undoubtedly important. Currently the following situation exists.
Despite the fact that combined $N$-body/hydrodynamic simulations show that the
method to estimate gravitational cluster masses from X-ray data (hydrostatic
assumption plus $\beta$ model) yields accurate and unbiased results at least out
to the radius within which the mean mass density equals 500 times the critical
density apart from major mergers (e.g., Schindler 1996; Evrard et
al.\ 1996), $M$--$T$ relations estimated from simulations and observations
disagree (but see the recent simulations by Thomas et al.\ 2002). This
disagreement needs to be solved.
\begin{figure}
%\vspace{-0.6cm}
\plottwo{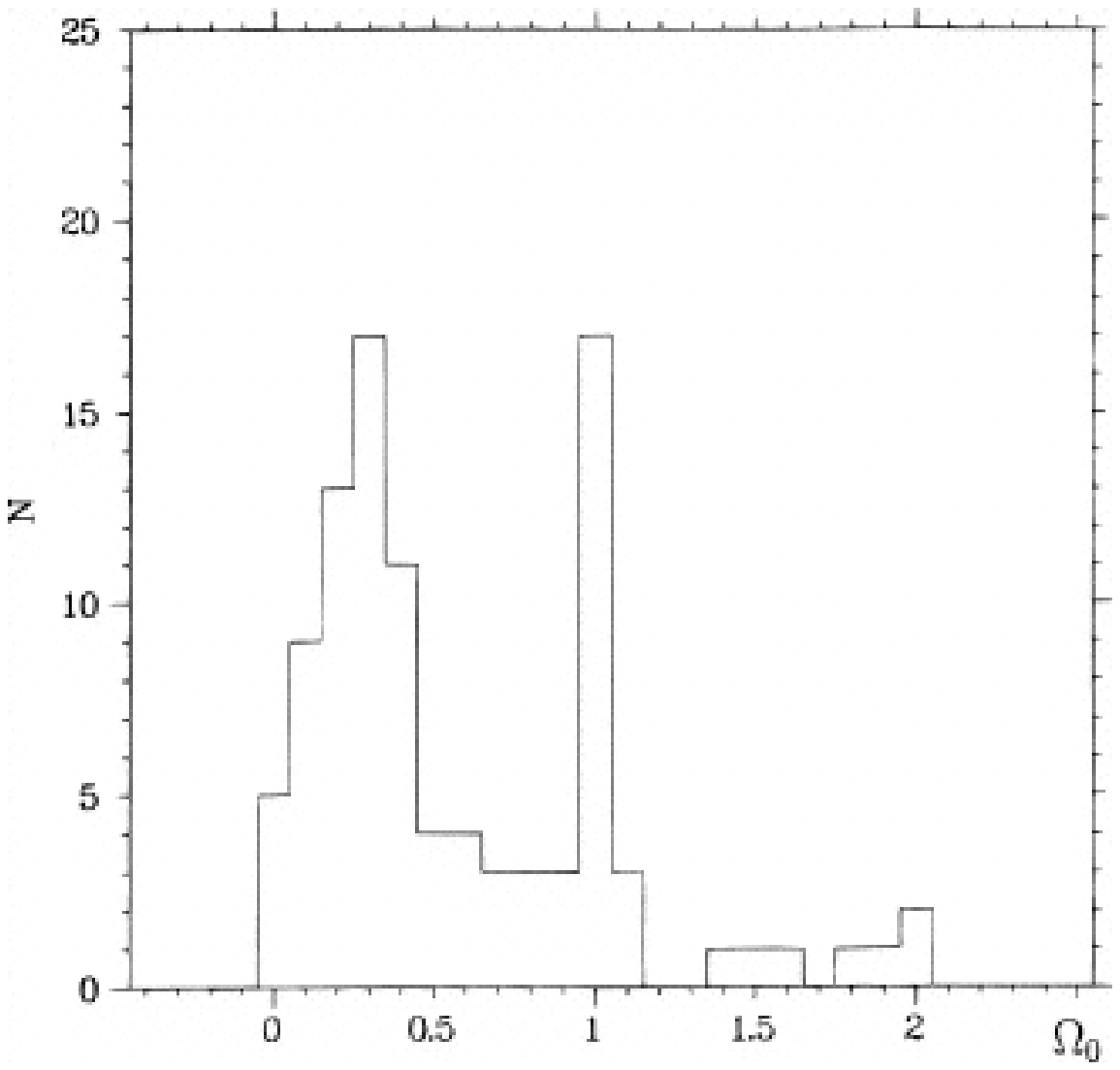}{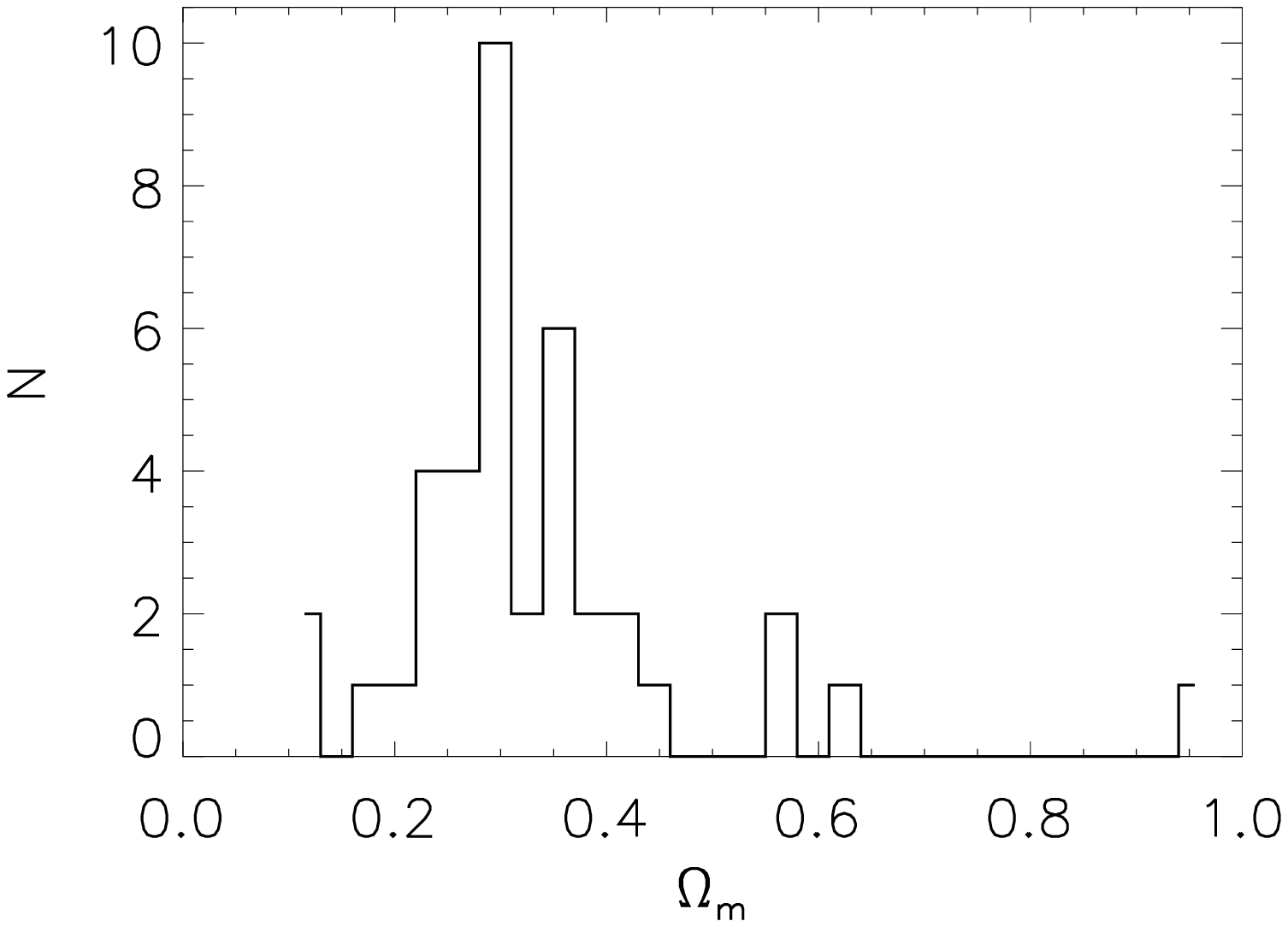}
%\vspace{-0.3cm}
\caption{{\bf Left:} Distribution of $\Omega_{\rm m}$ estimates compiled by Schuecker
et al.\ (1998). {\bf Right:} Distribution of $\Omega_{\rm m}$ estimates since 1997
compiled according to the criteria detailed in the text.}
%\vspace{-0.48cm}
\end{figure}

There are several possible ways how one may obtain the `correct' normalization
of the $M$--$T$ relation. The most direct way appears to be to study in more
detail simulations and observations.
From the simulation side one may test the effect of additional physics and
from the observational side clearly better spatially and spectrally resolved
X-ray observations are needed. We are currently in the most fortunate situation to
have two great X-ray observatories in space, XMM-Newton and Chandra, that
complement each other in the sense that they allow to study the pressure
structure in the outer cluster parts as well as in the very centers of clusters
in unprecedented detail. Observations of a reasonably sized cluster
sample and systematic comparison to mass estimates based on weak lensing and
galaxy velocity dispersions will allow to tighten the constraints on the $M$--$T$
relation significantly. Moreover such observations at hand for a complete
sample, like \gcs, one does not have to
resort to using a relation between temperature and mass but one can use the
determined masses directly for the construction of the mass function. Taking
into account the additional information on the gas density structure alleviates
some of the other systematic uncertainties discussed above.
It is obvious that such observations are a necessary step to ascertain that
galaxy clusters remain competitive 
cosmological probes needed for cross checks of systematic uncertainties inherent
to all measurements, to help break parameter degeneracies, e.g., for the
equation of state parameter, $w$, and to finally arrive at a detailed refined
theory of structure formation.

Let's return to discuss presently available constraints.
In Fig.~3 the distribution of a subset of published $\Omega_{\rm m}$ estimates
is shown in order to get a very rough overview of the observationally favored
values. 
The left hand side shows a compilation by Schuecker et al.\ (1998) and the right
hand side shows all published values that fufill the following criteria. Only
publications returned by a query of the Astrophysics Data System (ADS) with
```omega;m'' AND constraint' in the abstract and within the date range
`01-1997--06-2002' have been included. Out of the 156 retrieved publications only
abstracts have been checked for constraints
on $\Omega_{\rm m}$. Multiple occurences of the same data (e.g., as AAS
abstracts) as well as upper limits on $\Omega_{\rm m}$ have been excluded. If
there was a choice values for a flat Universe and $H_0=70\rm \,km\,s^{-1}Mpc^{-1}$
have been
used. If only an uncertainty range was given the mean value has been used. It is
quite obvious that this procedure excludes a large fraction of published
estimates on $\Omega_{\rm m}$ and certainly may be biased. Also error bars have been
entirely neglected. Still the resulting distribution might be useful to get a
feeling for the general trend.

A comparison of both histograms indicates that $\Omega_{\rm m}=1$ is less favored
nowadays than it used to be. Furthermore the mean value of the estimates today
appears to be lower than previously and especially values very close to
$\Omega_{\rm m}=0.3$ seem to be strongly supported. The best fit value
$\Omega_{\rm m}=0.12$ found from the \gcs\ mass function is therefore on the low
side of the current estimates, as expected from the left hand side of Fig.~2.
However, when plotting the estimates compiled here as a function of
time (Fig.~4) there appears to be a vague hint that the apparent decrease of the
mean might continue to the current date. Assuming boldly that improved accuracy
is positively correlated with publication date a `true' low value $\Omega_{\rm
m}\approx 0.2$ does currently not appear entirely unreasonable.
\begin{figure}
%\vspace{-0.6cm}
\plottwo{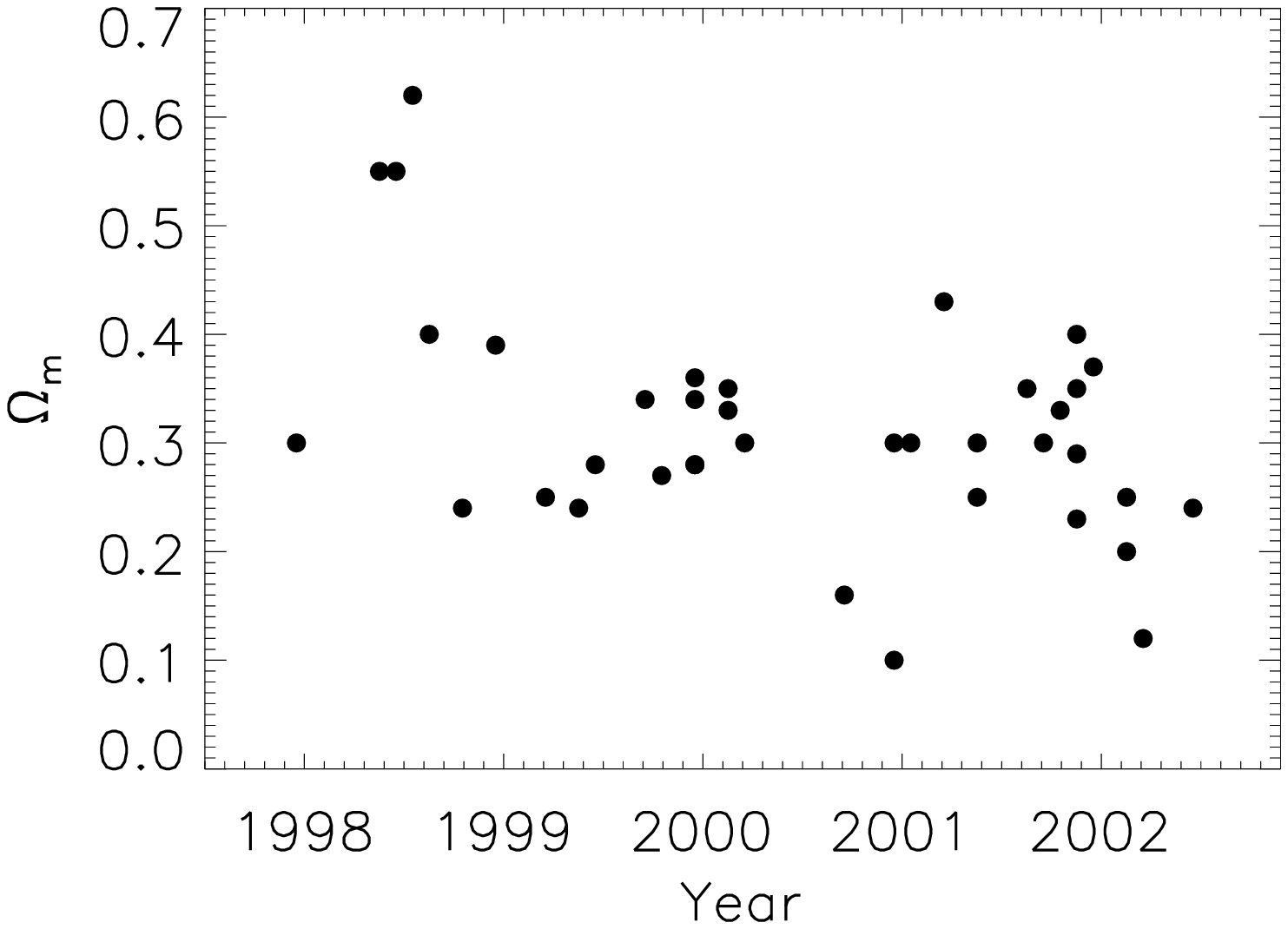}{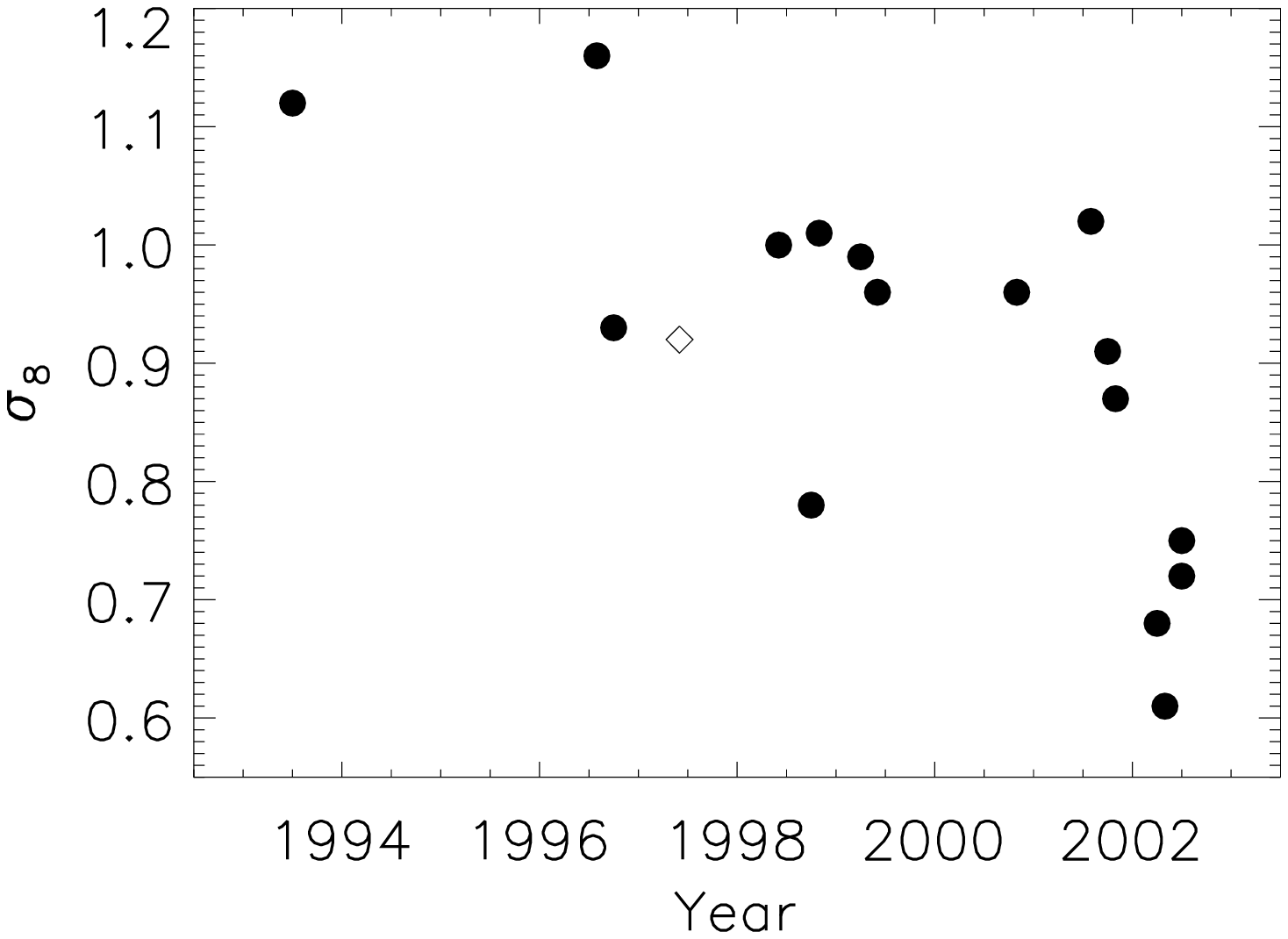}
\caption{{\bf Left:} Estimates of $\Omega_{\rm m}$ since 1997 (see text). Not
shown is an estimate of $\Omega_{\rm m}=0.96$ from 1999. {\bf
Right:} Estimates of $\sigma_8$ from cluster abundances for a fiducial value
$\Omega_{\rm m}=0.3$ as taken from the examples given in Reiprich \& B\"ohringer
(2002) plus three more recent estimates from Seljak (2002), Viana et al.\
(2002), and Bahcall et al.\ (2002). The diamond indicates the COBE normalization
for a flat Universe with $\Omega_{\rm m}=0.3$.} 
\end{figure}

The evolution of cluster abundances
is given by the
graph on the right hand side of Fig.~4 where the $\sigma_8$ estimates from
Fig.~2 are 
plotted for a fiducial value $\Omega_{\rm m}=0.3$ as a function of publication
date (results not yet published in a journal have been assigned the date
06-2002). Here one can make out an indication of decreasing $\sigma_8$
estimates with time and a cluster normalization substantially lower than
previously thought might be suggested (for the specific fiducial choice
$\Omega_{\rm m}=0.3$; for $\Omega_{\rm m}\approx 0.15$ a value close to one for
$\sigma_8$ may be recovered).
However, it cannot be stressed enough that the simple minded plots shown in this
section might be completely
misleading and should not be taken too seriously. By the time these conference
proceedings get published hopefully first results from MAP and especially their
combination with independent estimates will have narrowed down the range of
reasonable parameter values.

\acknowledgements
The code for calculation of the Bunn \& White normalization has been provided
by T. Kitayama.
T.H.R. thanks the Celerity Foundation for support. T.H.R. was        
also supported by NASA XMM-Newton Grant NAG5-10075.
For more infos on \gcs\ visit http://www.reiprich.net\,.
%\vspace{-0.6cm}

\end{document}